\documentclass[preprint,prd,noshowpacs]{revtex4}
\usepackage{amssymb}
\usepackage{amsmath}

\begin{document}
\title{A test of the circular Unruh effect using atomic electrons}
\author{Navid Rad}
\email{navidkrad@mail.fresnostate.edu}
\author{Douglas Singleton}
\email{dougs@csufresno.edu}

\affiliation{Physics Department, CSU Fresno, Fresno, CA 93740 USA}

\date{\today}

\begin{abstract}
We propose a test for the circular Unruh effect using certain atoms -- fluorine and oxygen. 
For these atoms the centripetal acceleration of the outer shell 
electrons implies an effective Unruh temperature in the range $1000 - 2000$ K.
This range of Unruh temperatures is large enough to shift the expected occupancy 
of the lowest energy level and nearby energy levels. In effect the Unruh temperature
changes the expected pure ground state, with all the electrons in the lowest energy level,
to a mixed state with some larger than expected occupancy of states near to the lowest energy level. 
Examining these atoms at low background temperatures and finding a larger than expected
number of electrons in low lying excited levels, beyond what is expected due to the background 
thermal excitation, would provide experimental evidence for the Unruh effect. 
\\
\end{abstract}

\maketitle

\section{Introduction} 

The Unruh effect \cite{unruh} - where an observer undergoing constant, linear acceleration detects 
a thermal flux of particles - is an important consequence of studying quantum fields 
in non-inertial frames of reference. In addition 
because of the equivalence principle the Unruh effect is 
connected with Hawking radiation \cite{hawking}. Thus finding experimental
evidence of the Unruh effect would provide indirect evidence for Hawking radiation. 
The Unruh temperature of the thermal spectrum measured by a {\it linearly} accelerating observer is
\begin{equation}
\label{t-unruh}
k_B T = \frac{\hbar a}{2 \pi c} ~,
\end{equation}
where $a$ is the acceleration of the observer and $k_B$ is Boltzmann's constant. From \eqref{t-unruh} it can be seen that 
this is a small effect for normal accelerations.  
To get a temperature of even 1 K one needs $a = 2.4 \times 10 ^{20} ~ m/s^2$. Generally such large accelerations 
are hard to obtain, especially for linear acceleration. However, for 
circular motion (e.g. electrons in the storage ring at LEP) it was 
noticed \cite{bell1} that one could  obtain centripetal accelerations $\approx 2.9 \times 10^{23} ~ m/s^2$. Using equation
\eqref{t-unruh} this would correspond to a temperature of $\approx 1200$ K. 
Such a temperature should lead to detectable consequences
and in fact it was claimed \cite{bell1} that this Unruh temperature would lead to
the electrons populating the upper energy level to a greater than expected degree. 
The electrons in the storage ring were taken to be in a magnetic field and thus had two energy levels -- a lower energy 
level with the electron spin anti-aligned with the magnetic field and an upper energy level
with the electron spin aligned with the magnetic field. The effective Unruh temperature of the electrons
in the storage ring would be large enough to ``thermally" excite some percentage of the electrons into
a higher energy level. This {\it observed} population of the upper energy level by electrons in storage rings 
is known as the Sokolov-Ternov effect \cite{sokter} and was 
also studied in \cite{jackson, derbenev, barber, akhmedov, sing}.
There are some questions \cite{unruh2} if this is really a confirmation of the Unruh effect.

Over the years there have been other proposals for experimentally confirming the
Hawking/Unruh effect either directly or via analog systems \cite{unruh1, schutzhold, belgiorno, rogers}. 
A recent review article \cite{matsas} contains references to further experimental proposals for testing the Unruh effect.

In this paper we point out the possibility of a new experimental test of the Unruh effect using the outer electrons 
of fluorine and oxygen. The electrons in these atoms 
experience large centripetal accelerations (${\cal O} (10^{23}) ~ m/s^2$) which are comparable to 
the centripetal accelerations of the electrons in storage rings studied in \cite{bell1}. 
Based on these accelerations the outer electrons of these atoms should experience large effective Unruh temperatures 
similar to those experienced by electrons in storage rings. One argument against the existence of this 
proposed effect is that unlike the electrons in storage rings, 
which follow classical paths, the electrons in atoms are in quantum eigenstates and naively can not be 
said to follow a classical path. However, our contention is that it is the acceleration which is important;
that even if this acceleration occurs for electrons in a quantum eigenstate it should lead to an effective Unruh temperature. 
To our knowledge there has not been any test to determine if a system undergoing acceleration 
at the quantum level experiences an Unruh temperature or not. Even if the effect we propose is not seen, this still
gives the interesting conclusion that there is then something fundamentally different 
about acceleration at the classical versus 
quantum level. One argument in favor of the existence of this 
effect is that in the path integral approach to quantum mechanics,
particles do follow or ``try out" every possible path, with each path weighted by 
$\exp[ i \cdot\rm{Action}]$. The classical path has the heaviest weight. 
As paths deviate more from this classical path their weight is less and they contribute less to the path integral. 
Our suggestion is that each path experiences some average acceleration (i.e. $a$) and associated with this
average acceleration there will be an effective Unruh temperature given by 
\begin{equation}
\label{t-gen}
k_B T = K \frac{\hbar a}{c} ~,
\end{equation}
where $K$ is a proportionality constant which depends on nature of the acceleration -- for linear acceleration 
$K=1/2 \pi$ from \eqref{t-unruh}; later we show that for circular acceleration $K=1/ 2 \sqrt{3}$ \eqref{t-unruh-cir}.
Since the temperatures from each of these paths can not cancel -- they can only add up -- the full path integral 
version of the Unruh temperature should then be of the form \eqref{t-gen} with some $K$ of order $1$.
The next ingredient needed to observe the effective 
Unruh temperature of \eqref{t-gen} is excited energy levels which are close enough to the lowest level. 
We predict that due to this Unruh temperature the pure, ground state with all the electrons in the lowest energy
levels will be shifted to a mixed state with some fraction of the electrons occupying energy levels
close to the lowest energy level. By performing measurements on these atoms and finding a larger than expected 
population of the low lying excited levels one could confirm the circular Unruh effect. Note this effect does
not change the value of the energy levels as is the case with the Lamb shift where the electrons interact with
their own quantized electromagnetic field, rather it changes a pure state (the electrons occupying only the 
lowest energy level in the limit when the ambient, external temperature is zero) to a mixed state (the
electrons, to some degree, occupying some nearby energy levels). The electrons in the atoms are interacting with
the external, quantized vacuum electromagnetic field and react as if exposed to a heat bath at temperature \eqref{t-gen}.   

\section{Review of the Unruh-DeWitt detector and its application to circular motion} 

In this section we make a brief review of an Unruh-DeWitt detector which is a device for measuring 
the excitation spectrum and temperature of a given space-time or space-time trajectory. We also provide some
mathematical details about why the path integral formulation of quantum mechanics justifies 
the application of the Unruh-DeWitt detector formalism to atomic electrons which are following ``quantum" paths. 

An Unruh-DeWitt detector is a quantum system with two energy levels, $E_0 <E$, with $E_0$ being the low level. By placing
an Unruh-DeWitt detector in some given space-time or allowing it to follow some space-time trajectory one can determine
if the space-time or space-time trajectory creates ``photons" with 
respect to the detector by seeing if the upper energy level, $E$,
is excited. By measuring the rate at which the upper level is excited 
one can extract a spectrum and effective temperature 
for the space-time or space-time trajectory. 

Following references \cite{akhmedov} \cite{BD} (which give more complete details of the construction of
an Unruh-DeWitt detector) we assume our quantum, two-level system is coupled to a scalar field, $\phi (x ^\mu (\tau))$
via a coupling $g \mu (\tau) \phi( x ^\mu (\tau))$. Here $g$ is a coupling 
constant, $\mu (\tau)$ is the detector's monopole moment, 
and $x^\mu (\tau)$ is the detector's trajectory as a function of its proper time, $\tau$.  Our ``photons" will
be quanta of the scalar field. Nothing is lost by this but one does not have to carry around the Lorentz indices
associated with a vector field quanta like a real photon. The transition rate per unit proper time, $T(E)$, for this
Unruh-DeWitt detector to be excited from the low energy $E_0$ to the high energy $E$ is 
\begin{eqnarray}
\label{t-rate}
T (E) &=& g^2 \sum _E | \langle E | \mu (0) | E_0 \rangle | ^2 \int _{-\infty} ^{+ \infty} 
e^{-i(E-E_0) \Delta \tau} G^+ (\Delta \tau) d(\Delta \tau )
\nonumber \\
&=& g^2 \sum _E | \langle E | \mu (0) | E_0 \rangle | ^2 {\cal F} (E) ~.
\end{eqnarray}
Here we have set $c=1$. $G^+ (x , x')= \langle 0 | \phi (x) \phi (x') | 0 \rangle$ 
is the positive frequency Wightman function
since we are interested in excitations from $E_0$ to $E$ rather than de-excitations 
from $E$ to $E_0$. Finally $\Delta \tau = \tau - \tau '$ is
the difference in proper times. In \eqref{t-rate} the part that is independent 
of the details of the detector (i.e. the coupling $g$, the values
of the energy levels $E_0$, $E$) is called the response function {\it per unit proper time}. 
The response function is \cite{BD}
\begin{equation}
\label{response}
{\cal F} (E)=\int _{-\infty} ^{+ \infty} e^{-i(E-E_0) \Delta \tau} G^+ (\Delta \tau) d (\Delta \tau).
\end{equation} 
The information about which particular space-time or space-time trajectory one is dealing with is embedded in $G^+ (x , x')$.
For a detector undergoing constant, linear acceleration $a$ 
the space-time trajectory is $x( \tau) = \left(\frac{1}{a}\sinh\left( a\, \tau \right), \,
\frac{1}{a}\cosh\left(a\, \tau \right), \, 0, \, 0\right)$ and the Wightman function for linear acceleration is
\begin{equation}
G^+ _{linear} (\Delta \tau) \propto \frac{a^2}{\sinh^2\left(\frac{a}{2}\, 
\left(\Delta \tau - {\rm i}\epsilon\right)\right)} ~. 
\end{equation}
This  Wightman function for linear acceleration leads to the response function 
\begin{equation}
{\cal F}_{linear} (E) \propto \left( 1 - exp\left( \frac{2 \pi \Delta E}{a \hbar} \right) \right)^{-1} ~,
\end{equation}
where $\Delta E = E - E_0$. One can see that this response function is Planckian and one can extract the temperature
$T_{linear} = a \hbar / 2 \pi k_B$. Of more relevance to classical ``electrons"
orbiting an atomic nucleus is circular motion with a trajectory $x(\tau) = \left(\gamma \tau,\, R\, \cos\left(\gamma
\, \omega_0 \, \tau\right), \, R\, \sin\left(\gamma \, \omega_0 \, \tau\right), \,0\right)$, 
where $\gamma = 1/\sqrt{1 - R^2 \, \omega_0^2}$
is the usual gamma factor, $R$ is the radius of the orbit and $\omega _0$ is the angular velocity. The centripetal
acceleration in the rest frame is $a=\gamma^2 v^2 / R$. The Wightman function for this trajectory is 
\begin{eqnarray}
\label{wightman-circular}
G ^+ _{circular} (\Delta \tau) & \propto & \frac{1}{ \gamma ^2 \,
\left(\Delta \tau - {\rm i}\, \epsilon\right)^2 - 4\, R^2 \,
\sin^2\left(\frac{\gamma \, \omega_0}{2}\, \Delta \tau\right)} \nonumber \\
& \approx & \frac{1}{\left(\Delta \tau - {\rm i} \, \epsilon\right)^2} \, \left(1 +
\frac{1}{12}\, \left(a\, \Delta \tau\right)^2 - \frac{1}{360 \, v^2}\,
\left(a\, \Delta \tau\right)^4 + \dots\right)^{-1} ~.
\end{eqnarray}
The expression in the second line is the expansion of $G ^+ _{circular}$ up to ${\cal O} (a^4 )$. If one expands
$G^+ _{linear}$ in a similar way one finds that up to ${\cal O} (a^2 )$ 
the Wightman functions for linear and circular motion are the same. 
In the relativistic limit \cite{bell1, akhmedov} one can take $G ^+ _{circular}$ to ${\cal O} (a^2 )$. Using this
approximation of $G ^+ _{circular}$ in \eqref{response} gives the following response function 
\begin{equation}
{\cal F}_{circular} (E)  \propto  a \, e^{- 2 \sqrt{3}\, \frac{\Delta E}{\hbar a}} ~.
\end{equation}
Although this is {\it not} a Planckian response function one can define an effective temperature from the exponent as
\begin{equation}
\label{t-unruh-cir}
k_B T_c \approx \frac{\hbar a}{2 \sqrt{3} c} ~.
\end{equation}
Unlike the case of linear acceleration it is not possible to find an analytic, 
closed form expression for ${\cal F}_{circular} (E)$. The circular response function
has been studied numerically \cite{muller} \cite{gutti} and these studies 
show that the circular response function although not
exactly Planckian has a Planckian-like form with an effective temperature given by \eqref{t-unruh-cir}. 

In the following section we want to apply the above analysis 
to atomic electrons. To this end we should give some justification 
for applying the results of a quantum two-level Unruh-DeWitt 
detector moving along a {\it classical} path to an electron in an
atomic orbital. The problem does not occur in replacing the 
Unruh-DeWitt detector by an electron since both are quantum systems
and in fact the electron, because of its spin, is a very good 
physical model for an Unruh-DeWitt detector as demonstrated
in \cite{bell1, akhmedov, sing}. The problem occurs in using a 
classical path to approximate an electron in a stationary
state orbital. The justification for this approximation comes from the 
path integral approach to quantum mechanics where the
motion of an quantum particle is given by the sum over all 
possible paths with each path weighted by $\exp[ i \cdot\rm{Action}]$.
The classical path contributes the most, but in principle every path makes 
a contribution, although the more a particular path 
deviates from the classical path the less it will contribute. Normally one 
does not use the path integral method to solve for
quantities like the orbitals of electrons in atoms since this 
is more easily accomplished by more elementary means. However, in \cite{ho} the hydrogen
atom was solved via the path integral approach. Thus in principle 
(and in some cases \cite{ho} in practice) one can treat
electrons moving around the atomic nucleus as moving along 
all possible paths but with the classical path contributing the most. 
It is in this sense that we will apply the results of an Unruh-DeWitt 
detector moving in a classical circular path to atomic electrons --
the classical path result is the lowest order approximation to 
the full path integral result. Note that the temperature of each possible
path will be positive definite thus it is not possible that the temperatures 
of the different paths cancel each other. Each path will
contribute some positive temperature to the full path integral so that
the expression in \eqref{t-unruh-cir} will be a approximate lower bound to the
exact temperature.
 
\section{Centrifugal acceleration in atoms} 

In \cite{bell1} it was the centripetal acceleration of electrons in storage rings
which was used in the Unruh temperature \eqref{t-unruh} to estimate the temperature. The centripetal acceleration
of these storage ring electrons, which are moving essentially at the speed of light, 
is given by $a = \gamma ^2 c^2 /R$ where $\gamma$ is the standard relativistic 
gamma-factor (which was ${\cal O}(10^5)$ for
the storage rings considered in \cite{bell1}), $R$ is the radius of the 
orbit, and $v \rightarrow c$ since the velocity of the electrons
was essentially the speed of light. As pointed out in \cite{bell1, akhmedov, sing} 
there were two important differences between between
the Unruh effect for linear acceleration and the Unruh effect for circular acceleration: (i) For circular motion the spectrum 
is not exactly thermal (ii) The temperature for the circular case did not have a simple expression as \eqref{t-unruh}. In the
relativistic limit one could define the effective temperature by \eqref{t-unruh-cir}. This expression 
implied an Unruh-like temperature for circular motion $\pi/\sqrt{3} \approx 1.8$ times that for linear acceleration. However,
since the spectrum is no longer thermal the definition of a temperature for 
circular acceleration was not as clear as for linear 
acceleration. In the present paper we will conservatively use the 
lower temperature \eqref{t-unruh} to estimate the temperature 
experienced by the outer electrons due to their centripetal acceleration. Using \eqref{t-unruh-cir} instead of \eqref{t-unruh}
would make the proposed effect even more pronounced. Also as mentioned in the previous section even \eqref{t-unruh-cir} is
a lower bound on the actual temperature one would get by doing the full path integral treatment of the problem. In the
full path integral solution each non-classical path would contribute some positive temperature so that the final result would
be larger than \eqref{t-unruh-cir}. 

The effect that we propose is that for certain atoms the centripetal acceleration of the outer shell
electrons is of order $10^{23} ~ m/s^2$. This is equal to or greater than the acceleration experienced by the 
electrons in the storage rings considered in \cite{bell1}. Thus we propose that the outer shell electrons for 
these atoms experience large effective temperatures in the range $1000 - 2000$ K. 
These atoms additionally have excited energy levels which are close enough to the lowest energy level 
so that this effective Unruh temperature can shift the expected ground state with all (or effectively all for
low ambient temperature) the electrons in the lowest energy level, to a mixed state where there is some
significant occupancy of the higher energy levels which are nearby the lowest energy level. 
The centripetal acceleration of an electron in an atom can be determined from the centripetal potential
\begin{equation}
\label{V-cen}
V_c (r) =  \frac{l(l+1) \hbar ^2}{2 m_e r^2}~,
\end{equation}
where $m_e$ is the electron mass. The centripetal force is determined 
via $F_c = -\nabla V_c (r) = - \partial V_c / \partial r$ 
and the centripetal acceleration is
\begin{equation}
\label{a-cen}
a_c (r) =  \frac{F_c}{m_e} = \frac{l(l+1) \hbar ^2}{m_e ^2 r^3}~.
\end{equation}
This is simply the centripetal acceleration calculated from the standard centripetal potential term in the
Schr{\"o}dinger equation with a radial potential. To avoid confusion 
we want to stress at this point that the Unruh-like radiation 
we will discuss, and which we claim can permanently shift some of the electrons from the lowest energy level
to nearby energy levels has nothing to do with the old classical arguments that a 
{\it classical} electron in orbit around an atomic nucleus 
would be unstable due to emission of {\it classical} radiation associated with its acceleration. 
For one thing, with hydrogen in the ground state, $l=0$ so from \eqref{a-cen} the centripetal 
acceleration would be zero and thus from \eqref{t-unruh} or \eqref{t-unruh-cir} our proposed effect would vanish. Also
our proposed effect involves the electrons {\it absorbing} radiation from the quantum vacuum {\it not} emitting radiation. 
The radiation and temperature we are discussing is purely quantum mechanical 
since in the limit $\hbar \rightarrow 0$ the acceleration and Unruh-like
temperature vanish and our effect goes away. In any case our proposed effect has nothing to do with the old arguments about the
instability of atoms due to {\it classical} radiation by the electrons. The effect we are proposing could be compared to the 
Lamb shift where an electron interacts with its own quantized electromagnetic 
field resulting in a tiny shift in energy levels. 
Our proposed effect comes from the electrons interacting with the background, quantized electromagnetic field rather than the
electrons self field. In this sense our proposed effect is similar to the 
Unruh effect, Hawking effect or Casimir effect all of which rely on quantized 
{\it background} electromagnetic fields. The Lamb shift is an ``internal" effect where the electron interacts with its
own quantized electromagnetic field, while the present proposed effect (as well as Hawking radiation, Unruh radiation, and the
Casimir effect) is an ``external" effect which is due to interaction with the background, vacuum fluctuations 
of the electromagnetic fields due to the acceleration of the system (an electron in our case). 
This difference is manifested in that for the Lamb shift the internal energy 
level structure of the system is altered whereas for
our proposed effect the internal energy levels of the system are not changed but the fraction of electrons occupying 
a particular level is shifted. The expected ground state where all, or almost all, of the electrons are in the lowest
energy level, is altered so that there is now predicted to be some occupancy of excited energy levels which are close to the
lowest energy level. This shift of the ground state occupancy is as if the atoms are sitting in 
a heat bath at some temperature given by \eqref{t-gen}.

Returning to \eqref{a-cen} and \eqref{t-unruh} (recall to be conservative we are estimating the temperature associated with
the acceleration via the linear formula \eqref{t-unruh} instead of the approximate circular formula \eqref{t-unruh-cir})
we note that in order to have a non-zero temperature, we need a non-zero acceleration which means we need $l \ge 1$.
For an outer electron with $l=1$, as is the case for fluorine and oxygen, and taking the radius to be the characteristic 
atomic size, $r=10^{-10} ~ m$, gives an acceleration and temperature of 
$a_c=2.68 \times 10^{22} ~ m/s^2$ and $T \approx 110 ~$K. While this
results is one order of magnitude less than the temperature found for circulating electrons in storage rings it is easy to 
see that, due the the inverse cubic dependence on the radius, as one makes the radius smaller than $r=10^{-10} ~ m$ one 
can get temperatures that are of the order or larger to those found for electrons in storage rings. Table \ref{atoms}
gives the radii and estimated Unruh temperature, from \eqref{t-unruh}, for oxygen and fluorine.
\begin{table}[htbp]
 \caption{Radius, centripetal acceleration and Unruh temperature of the outer shell electrons}
\begin{center}
\begin{tabular}{|p{3cm}|p{3cm}|p{4cm}|p{4cm}|}
\hline
Atom & radius \footnote{The radius is defined by the peak of the calculated charge density of the 
outer orbital \cite{gasiorowicz}} & centripetal acceleration  & Unruh temperature \\
\hline
Oxygen &	0.45 $\times 10 ^{-10} ~ m$ &	2.94 $\times 10 ^{23} ~ m/s^2$ & 1200 ~ K \\
Fluorine &	0.40 $\times 10 ^{-10} ~ m$  &	4.19 $\times 10 ^{23}  ~ m/s^2$ &	1700 ~ K \\
\hline
\end{tabular}
\end{center}
\label{atoms}
\end{table}
Both of these atoms have outer shell electrons with $l=1$. We did not consider helium 
which has the smallest radius because its outer electrons have $l=0$ and thus from \eqref{a-cen} have
$a_c =0$. Neon is also not a good candidate. Although its outer shell electrons have a smaller radius 
and therefore higher effective Unruh temperature than fluorine or oxygen, 
the low lying excited levels are too far above the lowest energy level for this temperature to be able to ``thermally"
excite a significant number of electrons into these excited levels. However, singly ionized neon (and also 
singly ionized nitrogen) do have low lying excited levels above the lowest energy level {\it and} their radius is smaller and 
thus the effective Unruh temperature is larger than for oxygen and fluorine. Thus these ions may also 
allow one to look for this effect. The difficulty with looking for this effect with neon or nitrogen ions
is that in order to keep them ionized they may need to be kept in a large background temperature. This would
mask the effect of the Unruh temperature. In this paper we focus on fluorine and oxygen, since 
for these neutral atoms one can test for the proposed effect with a low background temperature. 

Again, one might object that for the electrons in storage rings considered in \cite{bell1} the paths followed
are classical, while the ``paths" of the electrons in the outer shells of atoms are quantum  eigenstates.
However, these electrons in the outer shell of atoms do experience a centripetal acceleration \eqref{a-cen}
similar to the electrons following classical paths in storage rings. It is on this basis of having a  
similar magnitude of acceleration that we contend that the outer electrons in fluorine and oxygen should also experience
an effective Unruh temperature. As mentioned before, if the outer electrons do not experience an
effective Unruh temperature in connection with the centripetal acceleration \eqref{a-cen}
this in itself would be an interesting discovery. It would show that there is some fundamental
difference between classical acceleration and quantum acceleration in the sense that acceleration 
in a quantum eigenstate does not lead to an Unruh temperature. 

In the next section we make the prediction that the low lying excited energy levels of the atoms in table
\ref{atoms} will become populated to a much greater degree than would be expected by the background temperature.
If this effect is not seen then it indicates there is some fundamental difference between classical and quantum accelerations.
One can raise a general objection to the existence of this process. If this 
effect really occurs it seems one would be able to continually extract energy from the vacuum as follows: 
(i) The Unruh temperature experienced by the outer electrons would excite them 
into the low lying excited levels; (ii) eventually these
Unruh excited electrons would jump back to the lowest energy level emitting a photon which could be captured as usable energy;
(iii) the electron in the lowest energy level would eventually be excited back into the low lying excited level by the Unruh
temperature and the process would repeat allowing a continual extraction of energy via the emitted photons.
This argument would seem indicate the non-existence of this process. However, while it is generally accepted that a 
quantum system with a lowest energy level and an excited level 
(i.e. and Unruh-DeWitt detector) has some probability to transition to
the excited level when it undergoes accelerated motion, there is still a question whether this excited/thermalized 
Unruh-DeWitt detector will radiate. In \cite{ford} Ford and O'Connell study a 
mechanical oscillator moving in one spatial dimension 
which is coupled to a scalar field. The oscillator plays the role of the Unruh-DeWitt detector and the scalar field
plays the role of the photon. They investigate the expectation of the total flux of field energy, $\langle j(y,t) \rangle$, 
radiated by the oscillator at some arbitrary point $y$. They get the following general relationship \cite{ford}
\begin{equation}
\label{oscillator}
\langle j(y,t) \rangle = \langle j_0 (y,t) \rangle + \langle j_{dir} (y,t) \rangle + \langle j_{int} (y,t) \rangle ~.
\end{equation}
In this formula $\langle j _0 (y,t) \rangle$ is the flux of scalar field radiation in the absence of the oscillator. The
second term, $\langle j _{dir} (y,t) \rangle$, is the power radiated by the oscillator and corresponds to the 1D analog of 
the Larmor formula (Ford and O'Connell find that this term is proportional to the square of the {\it velocity} of the
oscillator rather than the square of the {\it acceleration} as is the case of the usual Larmor formula. This difference  
comes about because Ford and O'Connell are using a scalar rather than a vector field). The third term, 
$\langle j _{int} (y,t) \rangle$, represents the flux of energy from the field back to the oscillator. The condition for 
oscillator to be in equilibrium is $\langle j _{dir} (y,t) \rangle = -\langle j _{int} (y,t) \rangle$ i.e. the
total flux of energy is zero $\langle j  (y,t) \rangle = 0$. Since $\langle j _0 (y,t) \rangle$ is the energy flux of the
scalar field in the absence of the oscillator this should be zero and this is confirmed by direct calculation in \cite{ford}.  
By direct calculation \cite{ford} it is shown that when the oscillator undergoes uniform, constant acceleration (i.e. 
hyperbolic motion) that the condition $\langle j _{dir} (y,t) \rangle = -\langle j _{int} (y,t) \rangle$ is satisfied.
As a result the oscillator will move from the lowest energy level to a higher energy and remain there. The loss in energy 
of the oscillator due to the  $\langle j _{dir} (y,t) \rangle$ term is made up by the gain in energy of the oscillator
due to the  $\langle j _{int} (y,t) \rangle$ term. Applying the result of Ford and O'Connell to the proposed circular Unruh
effect on atomic electrons in fluorine and oxygen would imply that there is a {\it permanent}, larger than 
expected occupancy of the electrons in excited energy levels which are close to the lowest energy level. 
Our proposed experiment below would provide a test of the correctness (or not) of the claim in 
\cite{ford} -- that an Unruh-Dewitt detector (the outer shell electrons in fluorine and oxygen
in our case) will have particles excited to the upper energy level(s) to a greater than expected degree depending on the Unruh
temperature but will not radiate.      

\section{Low lying energy levels of fluorine and oxygen} 

In addition to the relatively large estimated  Unruh temperatures of fluorine and oxygen given in 
table \ref{atoms}, these atoms also have low lying excited levels -- $0.05$ eV or less above the lowest level.
The ground level of fluorine is $^2$P$_{3/2}$ and for oxygen is $^3$P$_2$. 
The energies and equivalent temperatures of the low lying excited levels of fluorine and oxygen
are given in table \ref{atoms2}.
\begin{table}[htbp]
\caption{Energy of the low lying, excited energy levels above the lowest level, 
equivalent temperatures, and spectroscopic notation.}
\begin{center}
\begin{tabular}{|p{3cm}|p{3cm}|p{4cm}|p{4cm}|}
\hline
Atom & spec. notation & $\Delta E_{i1} = E_i - E_1$ 
\footnote{The energies of the low lying levels, $i=2,3$, above the lowest level, 
$1$, are found at \cite{nist} where they are given
as inverse wavelengths in $cm ^{-1}$. Here we have converted the inverse 
wavelengths to energies in eV} & $T=\Delta E_{i1} / k_B$  \\
\hline
Oxygen &	$^3$P$_1$ ~;~ $^3$P$_0$ & 0.02 eV ~~;~~  0.03 eV & 232 K ~~;~~ 348 K\\
Fluorine &	$^2$P$_{1/2}$ &	0.05 eV & 580 K \\
\hline
\end{tabular}
\end{center}
\label{atoms2}
\end{table}
The prediction we make is that the estimated effective Unruh temperatures in table \ref{atoms}
will ``thermally" shift the occupancy of the lowest energy level and nearby energy levels 
so that the low lying excited levels will be populated by more electrons than would be
expected from the background temperature -- assuming that the background temperature is room temperature ($\approx 300$ K)
or less. To make the contrast bigger, we assume a background temperature of 100 K, which is above the 
boiling point for both gases. Thus the atoms would be in the gas phase. To calculate the 
fraction of the electrons populating the low lying excited levels listed in 
table \ref{atoms2} due to thermal excitation we use the density matrix formalism. The density matrix is
\begin{equation}
\label{density}
\rho = \frac{\exp (- {\hat H} / k_B T) }{Z} ~,
\end{equation}
where $Z=Tr(\exp(-{\hat H} /k_B T) = \sum _j \exp (-E_j /k_B T)$ is the partition function and ${\hat H}$ is the Hamiltonian
operator for the system. The fraction of electrons in a given low lying excited levels are given by
\begin{equation}
\label{f}
f(E_i, T) = \rho_{ii} = \frac{\exp(-E_i / k_B T) }{ \sum _j \exp (-E_j /k_B T)} ~.
\end{equation}
For oxygen and fluorine \eqref{f} becomes
\begin{eqnarray}
\label{fermi-dirac}
f_{Oxy} (E_{i=2,3}, T) &=& \frac{1}{\exp \left( \frac{\Delta E_{i1}}{k_b T} \right) 
+ \exp \left( \frac{ \pm \Delta E_{23}}{k_b T} \right)  + 1} ~, 
\nonumber \\
f_{Fl} (E_2, T) &=& \frac{1}{\exp \left( \frac{\Delta E_{21}}{k_b T} \right) + 1} ~.
\end{eqnarray}
These expressions assume that only the lowest energy level and the low lying excited level(s) are relevant. This is 
a good approximation for the effective Unruh temperatures or background temperatures that we deal with. In 
\eqref{fermi-dirac} $\Delta E_{23} =E_2 -E_3 = -0.01$ eV and $i=2$ goes with 
$+\Delta E_{23}$ and $i=3$ goes with $-\Delta E_{23}$.
Using \eqref{fermi-dirac} and the data from table \ref{atoms2} we give the expected fraction of electrons 
excited into the low lying excited levels for background temperatures $T=100$ K, for $T=300$ K and for 
the effective Unruh temperatures given in table \ref{atoms}. 
From table \ref{atoms3} one can see that the difference in the fraction of 
electrons populating the low lying excited levels based on the background temperature versus
the effective Unruh temperature can be significant especially if one does the measurement at low
background temperature (i.e. $T=100$ K). The difference between the fraction of electrons populating
the low lying excited levels due to the proposed Unruh effect versus what one would
expect due to room temperature ($T=300$ K) is not so big for oxygen; for fluorine the estimated 
Unruh excited fraction is only 3 times as large as the expected excitation due to room temperature.
This may explain why this effect has not been noticed so far since one needs to measure the population of the
low lying levels at the lowest possible temperature. Thus our prediction is that if one measures the
fraction of electrons populating these low lying energy levels for oxygen and fluorine, due to the 
effective Unruh temperature, these levels will be populated to a greater than expected degree,
especially if the comparison is made with low background temperature. One concrete way to test for this predicted 
higher occupancy of the low lying excited level(s) versus the lowest energy level would be to chose some higher
excited level and send in photons corresponding to the energy of transition between the lowest energy level and the higher 
excited level. Next send in photons corresponding to the energy of transition between the low lying excited level(s) and the
higher excited level. If the low lying excited levels are populated to a greater than expected degree due to
the proposed effect the absorption of the photon corresponding to the transition between the low lying excited level(s)
and the higher excited level should be proportionally larger. 
\begin{table}[htbp]
\caption{The fraction of electrons populating the low lying excited levels from table \ref{atoms2} assuming
these levels are populated by thermal excitations of a background temperature versus the effective Unruh temperature}
\begin{center}
\begin{tabular}{|p{2cm}|p{2.5cm}|p{3.5cm}|p{3.5cm}|p{3cm}|}
\hline
Atom & configuration & $f(T = 100 K)$  & $f(T = 300 K)$ & $f(T_{Unruh})$ \\
\hline
Oxygen &	$^3$P$_1$ ~~;~~ $^3$P$_0$ &	0.07 ~~;~~ 0.03 & 0.22 ~~;~~ 0.21 & 0.30 ~~;~~ 0.31 \\
Fluorine &	$^2$P$_{1/2}$ &	0.003 & 0.13 & 0.42 \\
\hline
\end{tabular}
\end{center}
\label{atoms3}
\end{table}
  
\section{Conclusion} 

In this paper we propose a possible test for the circular Unruh effect where the outer electrons 
of certain atoms (fluorine and oxygen) experience
large centripetal accelerations \eqref{a-cen} which should correspond to large effective Unruh temperatures \eqref{t-unruh}.
Although it is hard/impossible to talk about a classical path for electrons in quantum energy levels, in the path integral 
approach to quantum mechanics the electron tries out every possible path, with each 
path being weighted by $\exp[i \cdot {\rm Action}]$.
The classical path has the largest weight. Our contention is that each path will have a temperature associated with it
and the path integral averaged sum of all these temperatures for all paths can be estimated via 
a conservative lower bound of \eqref{t-unruh}. With the estimated Unruh temperatures given in table \ref{atoms}, all of 
which are larger than the assumed background temperature, occupancy of the electrons
in the lowest energy level and nearby levels will be ``thermally" shifted -- there will be a larger
than expected number of electrons occupying the low lying excited levels. The effect proposed here
for atomic electrons is similar to the Bell and Leinaas effect, except that the classical 
path of \cite{bell1} \cite{akhmedov} \cite{sing}
is replaced by quantum paths (i.e. a path integral). We stress again that 
this proposed effect has nothing to do with the old arguments 
for the instability of atoms due to classical radiation of electrons. In our proposed effect the occupancy of the 
electrons in the low lying excited level(s) are ``thermally" shifted by quanta from the vacuum. 
This is similar to the Lamb shift where hydrogen energy levels are shifted due to the interaction
of the electron with its own quantized ``internal" electromagnetic field. 
In the present case the electrons are interacting with the
``external" quantized vacuum electromagnetic fields. Another distinction is that 
in the present case the energy levels are not shifted but the ground state occupancy of the levels
is permanently altered. The permanence of this shift in the occupation of the low lying excited energy levels follows from 
the results of \cite{ford} -- although the electrons are shifted to low 
lying excited levels as if they were exposed to a temperature bath
they will not de-excite. This is the result of the balancing of the flow of energy from the oscillator to the field 
(i.e. $\langle j _{dir} (y,t) \rangle$) with the flow of energy from the field 
to the oscillator (i.e. $\langle j _{int} (y,t) \rangle$).
For hyperbolic motion this means $\langle j _{dir} (y,t) \rangle = -\langle j _{int} (y,t) \rangle$. The works \cite{grove}
\cite{raine} come to similar conclusions.   

There are other possibilities for testing for this effect. First, as briefly 
discussed above, one could look at singly ionized nitrogen or singly ionized neon. 
The radii for both ions are less than the radius of neutral neon (i.e. $r \le 0.35 \times 10 ^{-10}$~m) 
which implies a temperature greater than
$T \ge 2500$ K. Also, in contrast to neutral nitrogen and neon atoms, both ions have
excited levels ($\Delta E_{N^+} = 0.006$ eV,  $\Delta E_{N^+} = 0.016$ eV and $\Delta E _{Ne^+} = 0.10$ 
eV respectively) which are low enough to be
significantly populated by the effective Unruh temperature $T \ge 2500$ K. 
The difficulty is that in order to keep nitrogen and neon ionized one may need to have a large
background temperature to begin with which would mask the proposed effect.   
A second possibility for testing this effect would be to induce a level splitting 
using an external magnetic field (i.e. use the Zeeman effect).  
For example, applying an external magnetic field to fluorine would split the ground state $^2$P$_{3/2}$ 
into four different levels with $m_j=+\frac{3}{2}$ and $m_j=+\frac{1}{2}$ levels shifted upward. 
The lowest lying excited levels of fluorine  $^2$P$_{1/2}$ would split into 
two levels with the $m_j=-\frac{1}{2}$ level shifted
lower. Thus the $m_j=+\frac{3}{2}$ and $m_j=+\frac{1}{2}$ levels of the ground state $^2$P$_{3/2}$ 
and the $m_j=-\frac{1}{2}$ level of $^2$P$_{1/2}$ 
lowest lying excited levels would have a smaller energy difference than without the magnetic field, and the effective 
Unruh temperature of fluorine, $T=1700$ K, will more easily be able to 
populate the $m_j=-\frac{1}{2}$ level of $^2$P$_{1/2}$.   
We are currently investigating these extensions of the proposed effect.      
A third possibility for looking for this effect would be to look at 
nuclei. In the shell model of the nucleus the individual nucleons
occupy shells which have orbital angular momenta. The characteristic nuclear size is 
$10^{-15}$ m as opposed to the characteristic atomic size of
$10^{-10}$ m. This five order of magnitude difference in size would greatly enhance the centripetal acceleration 
of \eqref{a-cen}. However, nucleons are $\approx$ 2000 times heavier than 
electrons which would make the centripetal acceleration smaller
by this amount. Still the decrease in size would still be the larger effect due to the $r^{-3}$ 
dependence in \eqref{a-cen}. But nuclear energy 
levels are generally in the MeV range rather than the eV range of atomic physics. 
The goal would be to find some nucleus with a small radius, whose
nucleons had a non-zero orbital angular momentum and which had low lying excited energy levels above the lowest level.  

In conclusion we predict that the population of the low lying excited levels of fluorine and oxygen will have a
higher than expected occupation (compare the last column in table \ref{atoms3} 
with the two preceding columns). Seeing this larger 
than expected occupation would provide evidence to the Unruh effect. To make the contrast the largest, the experiment should
be done at the lowest possible background temperature. Not seeing this larger than expected occupation would indicate some
fundamental difference between acceleration at the classical versus quantum level.


\begin{thebibliography}{99}

\bibitem{unruh} W.G. Unruh, Phys. Rev. D {\bf 14}, 870 (1976).

\bibitem{hawking} S.W. Hawking, Comm. Math. Phys. {\bf 43}, 199 (1975).

\bibitem{bell1} J.S.Bell and J.M.Leinaas, Nucl.Phys. B {\bf 212}, 131 (1983); Nucl.Phys. B {\bf 284}, 488 (1987).

\bibitem{sokter} A.A.Sokolov and I.M.Ternov, 1963, Dokl. Akad. Nauk SSSR
{\bf 153}, 1052. 
 
\bibitem{jackson} J.D.Jackson, Rev.Mod.Phys. {\bf 48}, 417 (1976).

\bibitem{derbenev} Ya.S. Derbenev and A.M. Kondratenko, Zh. Exp. Theor. Phys. {\bf 64},1918 (1973).

\bibitem{barber} D.P. Barber and S.R. Mane, Phys. Rev. A, {\bf 37}, 456 (1988).

\bibitem{akhmedov} E.T. Akhmedov and D. Singleton, Int. J. Mod. Phys. A, {\bf 22} 4797 (2007).

\bibitem{sing} E.T. Akhmedov and D. Singleton,  Pisma Zh. Eksp. Teor. Fiz. {\bf 86}, 702 (2007). 

\bibitem{unruh2} W.G. Unruh, Phys. Rept. {\bf 307}, 163 (1998) and references therein.

\bibitem{unruh1} W. G. Unruh, Phys. Rev. Lett. {\bf 46}, 1351 (1981).

\bibitem{schutzhold} R. Sch{\"u}tzhold, G. Schaller, and D. Habs, Phys. Rev. Lett. 97, 121302 (2006).

\bibitem{belgiorno} F. Belgiorno et al, Phys. Rev. Lett. {\bf 105}, 203901 (2010).

\bibitem{rogers} J. Rogers, Phys. Rev. Lett. {\bf 61}, 2113 (1988).

\bibitem{matsas} L.C.B. Crispino, A. Higuchi, G. E.A. Matsas, Rev. Mod. Phys. {\bf 80}, 787 (2008).

\bibitem{BD} N.D. Birrell and P.C.W. Davies, ``Quantum fields in curved space",
(Cambridge University Press, Cambridge 1982).

\bibitem{muller} D. M{\"u}ller, ``A Semi-Analytical Method For The Evaluation of The Power Spectrum of a Rotating
Observer", arXiv:gr-qc/9512038.

\bibitem{gutti} S. Gutti, S. Kulkarni and L. Sriramkumar, Phys. Rev. D {\bf 83}, 064011 (2011).

\bibitem{ho} R. Ho and A. Inomata, Phys. Rev. Lett. {\bf 48}, 231 (1982).

\bibitem{ford} G.W. Ford and R.F. O'Connell, Phys. Lett. A {\bf 350}, 17 (2006). 

\bibitem{gasiorowicz}  S. Gasiorowicz, ``Quantum Physics", 2$^{nd}$ edition page 319 (John Wiley \& Sons, Inc. 1996).

\bibitem{nist} http://www.nist.gov/pml/data/handbook/index.cfm

\bibitem{grove} P.G. Grove, Class. Quant. Grav. {\bf 3}, 802 (1986).

\bibitem{raine} D.J. Raine, D.W. Sciama, and P.G. Grove, Proc. Roy. Soc.Lon. A {\bf 435}, 205 (1991).  


\end{thebibliography}
\end{document}